\documentclass[a4paper,11pt]{article}
\pdfoutput=1 

\usepackage{jinstpub} 

\usepackage{lineno}

\title{\boldmath The DUNE Vertical Drift Photon Detection System}

\author{L. Paulucci,\note{Corresponding author.}}
\affiliation{Universidade Federal do ABC,\\Av. dos Estados, 5001, Santo André, SP, 09210-170, Brazil}

\emailAdd{laura.paulucci@ufabc.edu.br}

\abstract{The Deep Underground Neutrino Experiment (DUNE) is a long-baseline neutrino experiment designed to mainly investigate oscillation parameters, supernova physics and proton decay. Its far detector will be composed of four liquid argon time projection chamber (LArTPC) underground modules, in South Dakota-USA, which will detect a neutrino beam produced at Fermilab, 1300 km away, where a near detector will be in place. The second DUNE far detector module, Vertical Drift, will be a single phase LArTPC with electron drift along the vertical axis with two volumes of 13.5 m x 6.5 m x 60 m dimensions separated by a cathode plane. The charge collection will be performed by two anode planes placed at the top and bottom of the module, each composed by stacked layers of a perforated PCB technology with electrode strips. The photon detection system (PDS) will make use of large size X-Arapuca tiles distributed over three detection planes. One plane will consist of a horizontal arrangement of double sided tiles installed on the high voltage cathode plane and two vertical planes, each placed on the longest cryostat membrane walls. A light active coverage of 14.8\% over the cathode and 7.4\% over the laterals should allow improvements in the low energy physics range that can be probed in DUNE, especially regarding supernova neutrinos ($\rm \sim10~MeV$). We present the initial characterization of the Vertical Drift PDS using a Monte Carlo simulation and preliminary studies on its reconstruction capabilities at the MeV scale. The information obtained with the PDS alone should allow determination of a neutrino interaction region with a precision of at least 65 cm for events with deposited energy above 5 MeV and the deposited energy can be reconstructed with precision better than 10\%, both at the center of the volume.}

\keywords{Noble liquid detectors (scintillation), Large detector-systems performance, Time projection chambers, neutrino detectors}

\collaboration[c]{on behalf of DUNE collaboration}

\proceeding{LIGHT DETECTION IN NOBLE ELEMENTS (LIDINE 2021)\\
  September 14th - 17th, 2021\\
  University of California San Diego, San Diego, USA}

\begin{document}
\maketitle
\flushbottom

\section{Introduction}
\label{sec:intro}

The Deep Underground Neutrino Experiment (DUNE) \cite{dune} is a long-baseline neutrino experiment, currently in the design phase and early construction activities, with main goals to study neutrino oscillation parameters, mass hierarchy, CP violation, proton decay, and supernova neutrinos. Its near detector and neutrino beam source will be located at the Fermi National Accelerator Laboratory (Fermilab), USA, while the far detector will be placed in the Sanford Underground Research Facility (SURF), about 1300 km away from Fermilab.
The far detector will be composed of four liquid argon time projection chambers (LArTPCs), each with a large fiducial volume ($\sim$10 kt), placed about 1.5 km underground.

In a LArTPC, charge generated by ionization is drifted 
towards a set of grids which allows the reconstruction of particles' trajectories inside the chamber. Argon scintillation light is also collected, complementing the ionization charge signal, providing fast timing information on the interactions and therefore being used in event time reconstruction, event triggering, and event calorimetry. 

The second DUNE far detector module, called Vertical Drift (FD2-VD), will be a LArTPC of size 13.5 m x 13.0 m x 60 m divided into two volumes by a central horizontal cathode. Each volume will have its own anode plane, parallel to the cathode, where electrons will be collected by circuit printed boards (CRPs). This module photon detection system (PDS) will use large 60 x 60 cm$^2$ X-Arapucas \cite{xarapuca}, a box with highly reflective internal walls and with a set of wavelenght shifters and a dichroic filter designed to trap photons on the inside of the device so they can be detected by silicon photomultipliers (SiPM) nested in the box laterals. They will be distributed over the cathode (X-Z plane), with a coverage of 14.8\%, and over the two laterals along the Y-Z plane, on the cryostat wall, with a coverage of 7.4\%. This layout does not accommodate photon detectors (PDs) behind the anode plane, as in traditional LArTPCs, due to the opacity to light of the PCB plane structure, and introduces the need to operate the cathode PDs on high voltage surfaces. To meet this challenging constraint, the SiPMs in cathode-mounted modules will be powered using power-over-fiber and read with signal-over-fiber technology, thus providing the desired voltage isolation.

The proposed PDS for the second DUNE far detector aims at providing a light yield approximately uniform over the whole detection volume. This solution features position reconstruction capability, in addition to an improved energy reconstruction resolution and lower detection energy threshold, which should allow for a broadening of the physics scope that this module can provide, in particular to the low energy physics aimed by DUNE. Here we present a first estimate on the FD2-VD PDS capabilities based on Monte Carlo (MC) simulations.

\section{Description of the simulation}

\begin{figure}[htbp]
\centering 
\includegraphics[width=.7\textwidth]{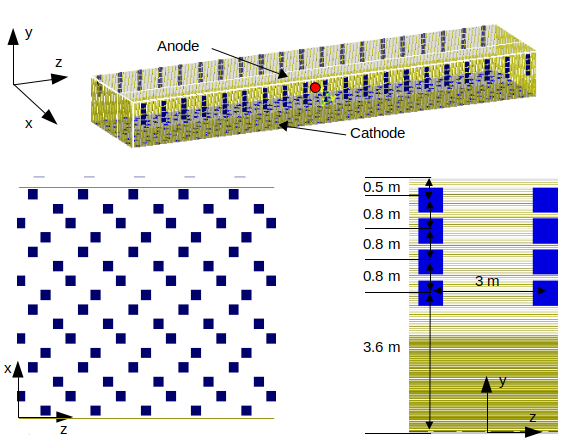}
\caption{\label{fig:sim} Simulated volume (top half of total volume where electron drift is in the positive y-direction) used in the physics studies. PDs are depicted in blue, while the field cage profiles are shown in yellow. On top, the full simulation volume where the red dot marks the origin of the coordinate system, on the bottom left, details on the cathode PD disposition while on the right, PD details over the cryostat walls.}
\end{figure}

A simulation of the FD2-VD photon response based on a simplified detector geometry composed of the top half of the full volume (mirror image of the bottom half), shown in figure~\ref{fig:sim}, has been carried out using the Geant4 \cite{geant4, geant4_2} platform to provide first indications on the expected PDS performance. Scintillation photons are sampled isotropically from a point of origin and transported through LAr, being subjected to Rayleigh scattering, absorption by residual impurities and detector materials, reflection on detector materials (on the anode plane or field cage), and detection on photo-sensitive X- Arapuca. Input detector parameters and photon emission/propagation constants used in the MC generation are listed in table~\ref{tab:sim}. The field cage aluminum profiles will need to be adapted to allow for a larger light transmission (narrower profile with 70\% forward transmission) in about 60\% of its length over the laterals along the Y-Z plane.

\begin{table}[htbp]
\centering
\caption{\label{tab:sim} Simulation input parameters and physical constants.}
\smallskip
\begin{tabular}{|l|c|c|}
\hline
{\bf Parameter} & {\bf Value} & {\bf Comment} \\ \hline
LAr photon yield & 25,000 ph/MeV &  25\% Singlet (fast) emission \\ 	(mip, @500 V/cm) &               &  75\% Triplet (slow) emission \cite{Doke, Hitachi} \\ \hline
Xe doping in Ar  &    $\mathcal{O}$(10 ppm)   & Light from Ar triplet decay fully transfers to Xe \cite{Wahl, Kubota}  \\ \hline
Rayleigh scattering    & $\lambda_R = 1$m & 128\,nm Ar light   \\
    length              &  $\lambda_R = 8.5$m & 176\,nm Xe light \cite{Babicz} \\ \hline
Absorption length               & $\lambda_{Abs} = 50$m & From $N_2$ molecules \cite{Wu} \\ \hline
Number of PDs      & 320  & 40 rows ($z$) of 8 PDs on cathode  \\
&  160   & 20+20 rows ($z$) of 4 PDs on cryostat walls \\ \hline
Detection efficiency    & $\epsilon_D=3\%$ & X-Arapuca @ 128nm and 176nm  \cite{Henrique, Brizzolari, Paulucci} \\ \hline
Field cage reflectivity & 26\% @176nm & More transparent in $\sim$60\% along the Y-Z plane\\ \hline
Anode reflectivity & 20\% @176nm& Averaged over CRPs' reflective and opaque portions \\\hline
\end{tabular}
\end{table}

\section{Results}

\begin{figure}[htbp]
\centering
\includegraphics[width=.49\textwidth]{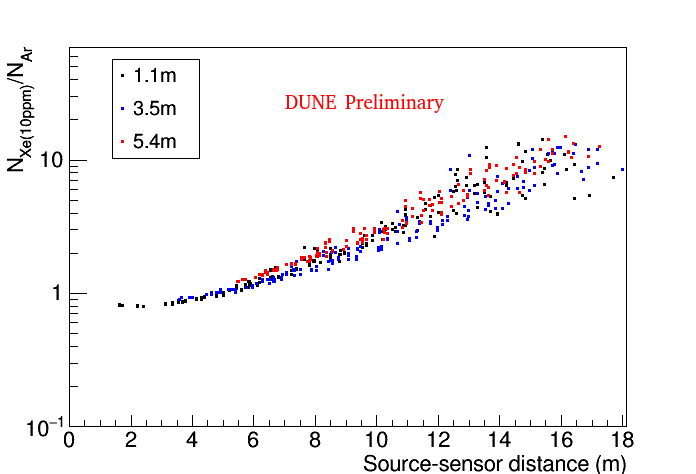}
\includegraphics[width=.49\textwidth]{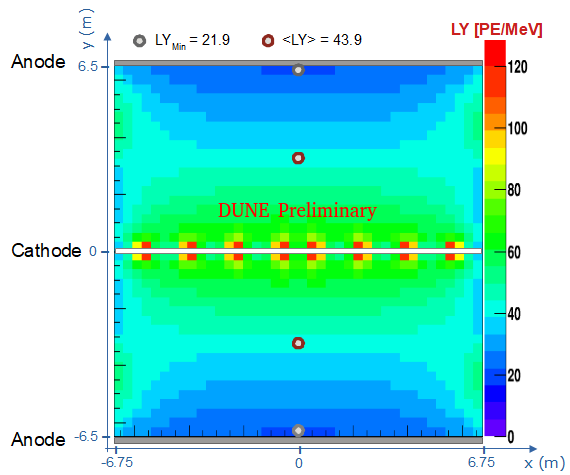}
\caption{\label{fig:xedoping} Left: Ratio of the number of photons detected per PD for Xe doped argon with respect to pure argon as a function of the distance from the light source. Different colors represent different heights of source with respect to cathode. Right: Average light yield map over X-Y for the central region of the detector.}
\end{figure}

{\bf Effects of Xe doping:} A first MC study has been performed to evaluate the effect of Xe doping on light collection by comparing the amount of photons reaching the PDs as a function of distance for LAr with and without Xe doping. The number of photons arriving at the i-th PD, at distance $d_i$ from the source, is obtained and sampled according to the X-Arapuca detection efficiency and the ratio between the two situations is seen in figure~\ref{fig:xedoping}, left. The collected light is found to be $\sim$23\% larger for Xe-doped Argon due to the effect of the longer Rayleigh scattering length enhancing collection probability for light emitted at longer distances from the detector. This supports the choice of Xe-doped Ar as scintillation medium for FD2-VD. 

{\bf Light yield:} The light yield map for the central transverse portion of the volume is given in figure~\ref{fig:xedoping}, right, averaged over 1.5 m (half the distance between lateral PD columns). The minimum LY is found for light emitted near the center of the anode plane, at the farthest distance from the photo-sensitive areas on cathode and lateral walls. The yield is sufficiently high ($LY_{min}$ > 20 PE/MeV and $\langle LY \rangle$ > 40 PE/MeV) to enable good detection efficiency for supernova neutrinos and proton decay searches anywhere in the fiducial volume. A lower LY is expected for the regions closer to either ends of the detector volume due to border effects. Thanks to the coverage over the laterals along the Y-Z plane, the LY non-uniformity along the drift (y) is reduced when compared to the PD response of having a coverage over only one plane.

\begin{figure}[htbp]
\centering 
\includegraphics[width=.49\textwidth]{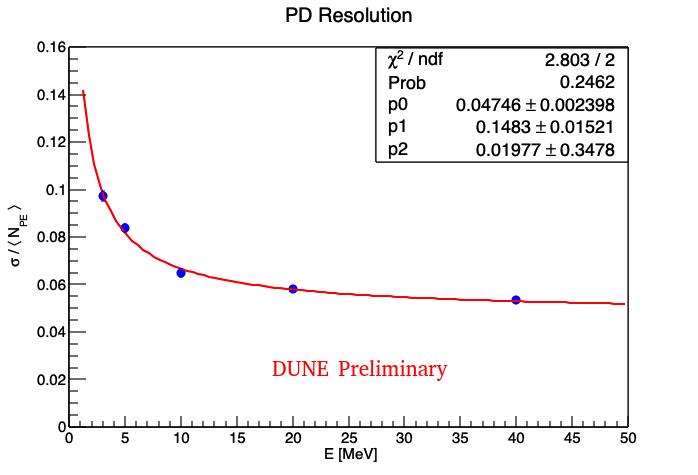}
\includegraphics[width=.49\textwidth]{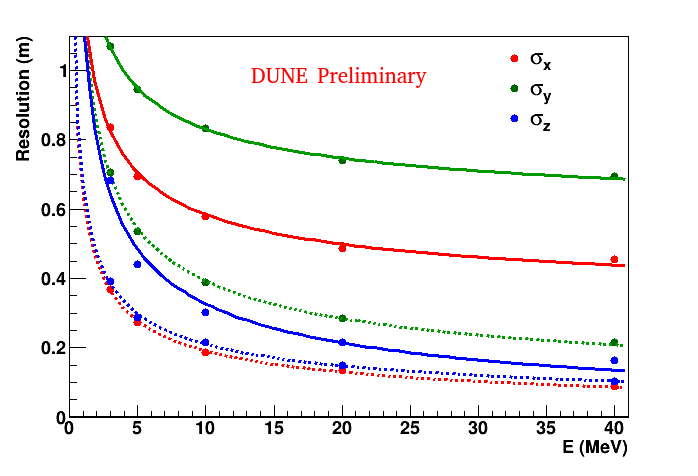}
\caption{\label{fig:EPosRes} Left: Energy resolution for low energy deposits at position (0,0,0) from PD signal collection. The resolution is fitted (red line) by the quadratic sum of constant ($p_0$), stochastic ($p_1/E$), and noise ($p_2/E$) terms. Right: Spatial resolution 
based on the standard deviation of the difference between reconstructed and true position coordinates. Full lines are for points in the central region ($\rm -20~m < z < 20~m$) 
of the FD2-VD volume while dashed lines are for position reconstruction performed for the central point (0,0,0).}
\end{figure}

{\bf Energy resolution:} The energy resolution at position (0,0,0), the center of the simulated FD2-VD volume, where LY$\sim$ 40 PE/MeV, has been evaluated by generating different energy deposits, recording photons that reach the PDs and simulating PE response from all hits. The expected energy resolution is shown in figure~\ref{fig:EPosRes}, left, and is better than 10\% for energy deposits larger than 3 MeV, being mainly affected by statistical fluctuations on the number of PEs detected and uncertainty on energy calibration. A 5\% systematic uncertainty on PE-to-energy calibration was adopted from the energy spread of the source (beam electrons) used in the latest Arapuca calibration test \cite{protodune}. Similar assumptions are expected for the DUNE experiment calibrations.

{\bf Spatial resolution:} Spatial event reconstruction capability with PDS signals alone should allow fast discrimination of background due to intrinsic radioactivity and external radiation from low-energy neutrino signals. The MC generation has been performed for point-like energy deposits and for each event (x, y, z, E), the number of detected PEs by each PD is obtained from simulation. An algorithm was developed for position reconstruction based on the barycenter of the light pattern of the event. The 
spatial resolution of the PD system is evaluated as the standard deviation of the differences between reconstructed and true positions for the simulated event sample. In figure \ref{fig:EPosRes}, right, the behavior of spatial resolution along the 3 coordinates with respect to the energy deposition, approximately inversely proportional to the square root of the number of photons detected in the event, is shown. As this number is proportional to the energy deposition, the precision of the reconstructed position improves for higher energy deposits. The position along the vertical coordinate is less precisely reconstructed because the optical coverage and average LY over the laterals are lower than over the cathode. In the x-direction, border effects decrease the resolution with respect to the z-direction while at the very center of the volume, these resolutions are comparable. Combination of light and charge signals should allow improvements on the position reconstruction, specially in the drift direction \cite{TDR}. Position determination by the PDS is also of particular interest in the event that the DUNE TPC cannot acquire charge, e.g., downtime of the TPC. 

\begin{figure}[htbp]
\centering 
\includegraphics[width=.49\textwidth]{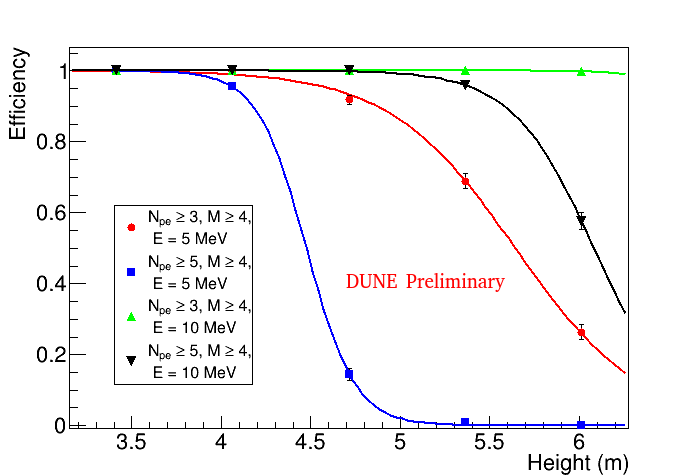}
\includegraphics[width=.49\textwidth]{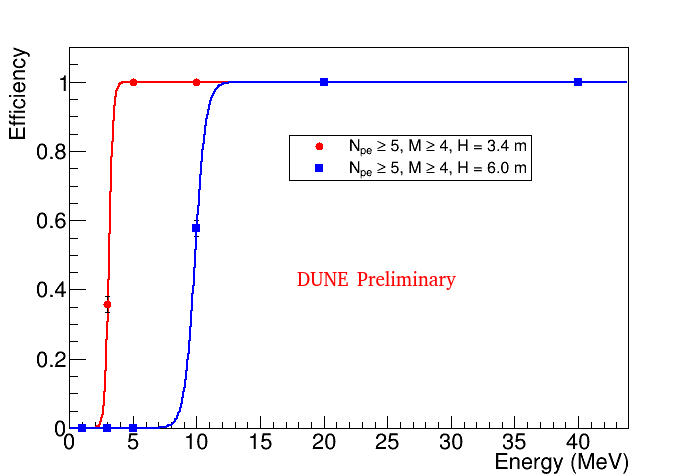}
\caption{\label{fig:trigger} Trigger efficiency in various trigger configurations.}
\end{figure}

{\bf Trigger efficiency for low-energy events:} The expected main feature of the enhanced FD2-VD PDS is a high trigger efficiency (ratio of events of interest selected-over-generated) down to low detection thresholds for low-energy astrophysical events. 
A trigger is fired when at least M adjacent PDs have crossed a N PE threshold (simple ‘MAJORITY OR’ trigger condition). The simulation was performed by generating low energy deposits at the region with lower average LY and larger LY gradient, i.e., at different heights at the center of the FD2-VD volume, along the vertical drift direction. Results are shown in figure~\ref{fig:trigger}. The trigger efficiency with (M=5,N=4)-Majority trigger is $\sim$100\% for $\geq$ 5 MeV deposits anywhere in the LAr volume up to y $\sim$ 4m (60\% of the drift distance). In the remaining 40\% of the volume, closer to the anode plane where the LY is minimal, full trigger efficiency is recorded for $\geq$ 15 MeV energy deposits. A more relaxed condition (for example, by lowering the number of PDs, M, and/or the threshold for PE detection, N) in this portion of the volume can still be applied, resulting in higher trigger efficiency for lower energy deposits at the cost of an increase in the rate of false-positive triggers, yet to be evaluated, due to radioactive backgrounds such as $^{39}$Ar with a 1.01 Bq/kg rate.

\section{Conclusions}

The most stringent requirements on the PDS performance in terms of light yield and timing resolution are imposed by the detection of nucleon decays and supernova neutrinos.  
We have shown with a Geant4-based MC simulation that the PDS proposed for the second DUNE far-detector module, Vertical Drift,  has the desired capabilities in terms of light yield (minumim above 20 PE/MeV and average above 40 PE/MeV), uniformity for light detection (with reduced dependence on the drift direction due to the PDs placed over the laterals along the Y-Z plane), energy (better than 10\%) and position (better than 1m) resolutions for low-energy deposits, and trigger capabilities (with expanded fiducial volume for full trigger at energy deposits of order of tens of MeV).
Moreover, it is foreseen the possibility of 
enhancing the physics reach beyond the minimal DUNE scientific requirements: triggering on galactic supernova burst events, determination of the drift position of proton decay signals and correction for charge lost due to electron capture and other transport effects in the TPC.
Thanks to an optimized PDS, the fiducial volume can be enlarged by increasing the capability to distinguish the signal from the external backgrounds entering the TPC volume. 
In the event that the DUNE TPC temporarily cannot operate at its goal electric field, the PDS energy measurements could compensate for reduced charge detector performance since light production increases relative to the free charge for lower electric field. 
It is important to point out that there is a strong R\&D program supporting the design choices of this module, in particular, the use of power- and signal-over-fiber technologies in LAr (see, for example, S. Sacerdoti's contribution to this volume).




\end{document}